\documentclass[a4paper]{ESASPCS13Style}
\usepackage{epsfig}

\begin{document}

\title{3D Simulation of Convection and Spectral 
       Line Formation in A-type Stars}

\author{M. Steffen\inst{1} \and B. Freytag\inst{2} \and H.-G. Ludwig\inst{3}} 
\institute{Astrophysikalisches Institut Potsdam, An der Sternwarte 16,
           D-14482 Potsdam, Germany \and 
           GRAAL, Universit{\'e} de Montpellier II,
           F-34095 Montpellier, France \and
           Lund Observatory, Box 43, 
           S-22100 Lund, Sweden}

\maketitle 

\begin{abstract}
  We present first realistic numerical simulations of 3D radiative
  convection in the surface layers of main sequence A-type
  stars with $T_{\rm eff} = 8000$~K and $8500$~K, $\log g = 4.4$ and $4.0$, 
  recently performed with the CO$^5$BOLD radiation hydrodynamics code. 
  The resulting models are used to investigate the structure of the 
  H+He\,I and the He\,II convection zones in comparison with the predictions
  of local and  non-local convection theories, and to determine the
  amount of `overshoot' into the stable layers below the He\,II
  convection zone. The simulations also predict how the topology of
  the photospheric granulation pattern changes from solar to
  A-type star convection. The influence of the photospheric temperature 
  fluctuations and velocity fields on the shape of
  spectral lines is demonstrated by computing synthetic line
  profiles and line bisectors for some representative examples,
  allowing us to confront the 3D model results with observations.

\end{abstract}

\begin{figure}[htb]
  \begin{center}
  \mbox{\includegraphics[width=9.0cm, clip=true]{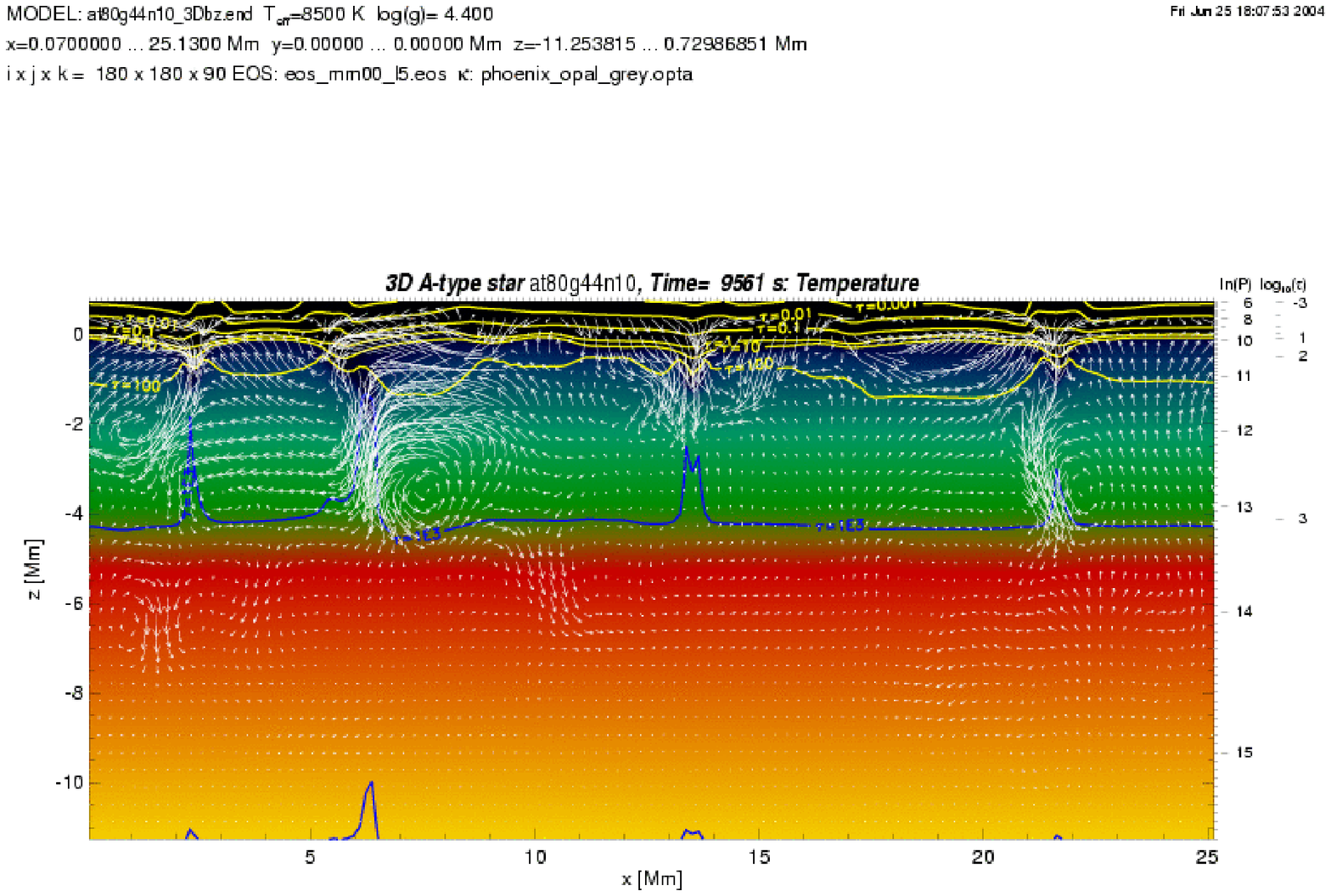}}
  \mbox{\includegraphics[width=9.0cm, clip=true]{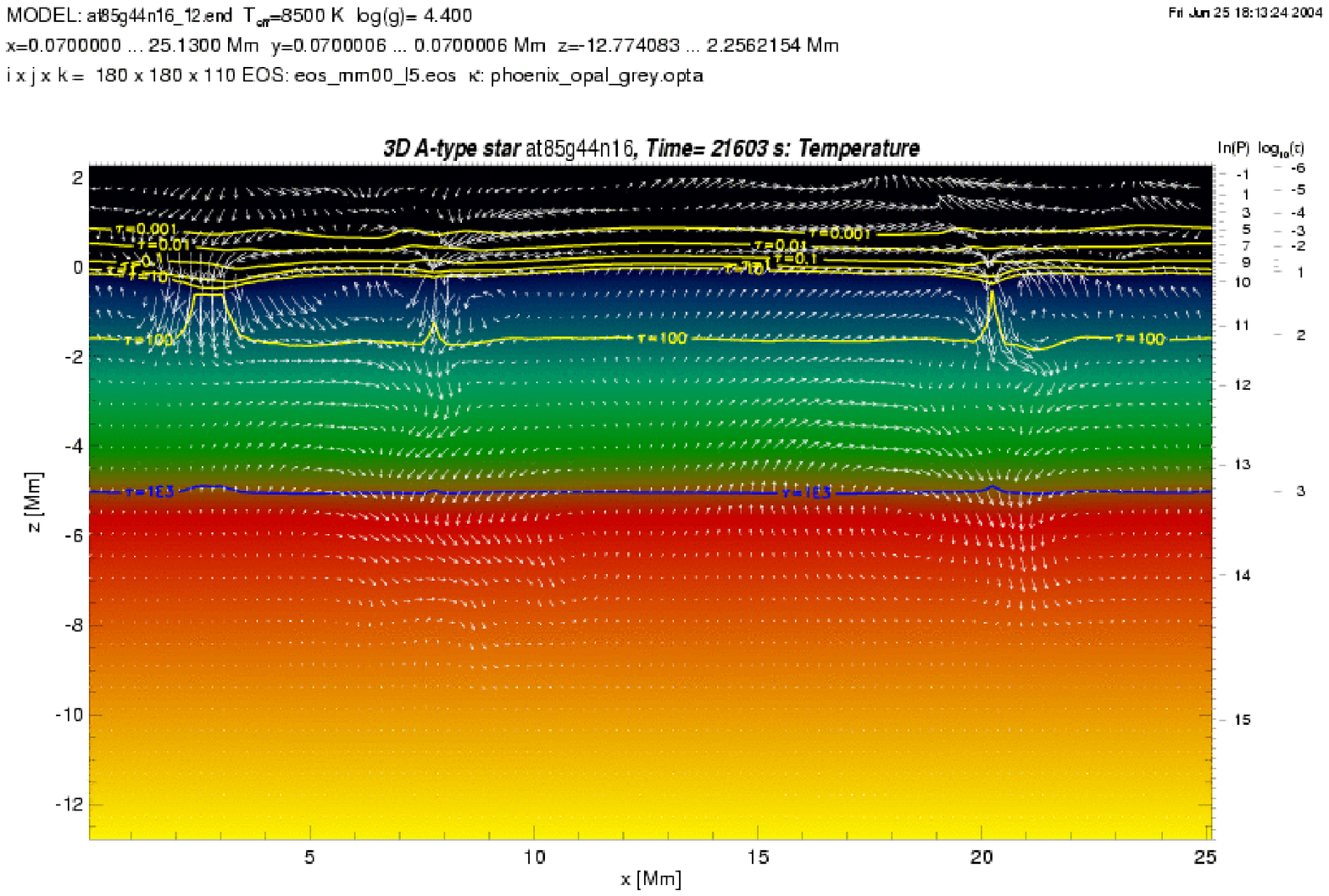}}
\end{center}
\caption{
    Arbitrary snapshots from two 3D simulations of \mbox{convection}
    in the surface layers of A-type stars, showing the 
    velocity field and temperature distribution in a vertical
    slice. {\bf Top:} \emph{MODEL~1:} $T_{\rm eff}=8000$~K, $\log g=4.4$, 
    geometrical size $25.2 \times 25.2 \times 12.0$~{\upshape Mm},
    $180 \times 180 \times 90$ grid cells, vertical optical depth
    range $-3 \le \log\tau_{\rm Ross} \le 4$, covering $\approx 10$~
    pressure scale heights.
    {\bf Bottom:} \emph{MODEL~2:} $T_{\rm eff}=8500$~K, $\log g=4.4$, 
    geometrical size $25.2 \times 25.2 \times 15.0$~{\upshape Mm} ,
    $180 \times 180 \times 110$ grid cells, 
    $-6 \le \log\tau_{\rm Ross} \le 4$, covering $\approx 16\, H_p$.
\label{fig1}}
\end{figure}

\section{Introduction}
  
In comparison with the Sun, convection in the envelopes of A-type stars
is a rather inefficient energy transport mechanism. According to local
mixing-length theory (MLT, \cite{bv58}) convection is confined to two 
separate shallow convection zones near the stellar surface. Unfortunately,
the structure and convective efficiency of these layers depends sensitively
on the choice of the --unknown-- mixing-length parameter. Another problem
with MLT is that it cannot describe convective overshoot. To overcome these
difficulties, 'parameter-free', non-local convection theories have been
developed and applied to A-type stars (\cite{km02}, KM02).

Radiation hydrodynamics simulations of stellar convection constitute
an independent approach. Up to now, \emph{realistic} simulations of
surface convection in A-type stars have been restricted to 2D
(\cite{fls96}).  3D simulations are challenging, because the short
radiative time scales in the atmospheres of these stars enforce an
exceedingly small numerical time step, and hence make convection
simulations for A-type stars much more time consuming than for the
Sun. On the other hand, A-type stars have the advantage that the 
entire convective part of the envelope can be included in a single 
simulation box with a simple closed lower boundary.

In the following, we present first results of 3D hydrodynamical
convection simulations for main-sequence A-type stars ($T_{\rm eff} =
8000$~K and $8500$~K, $\log g = 4.4$ and $4.0$, solar metallicity),
and compare them with the aforementioned convection theories. We also
address the interesting question of whether the hydrodynamical models
can reproduce the peculiar line profiles and lines asymmetries
(inverse bisector C-shape) observed for slowly rotating A-type stars
with $T_{\rm eff}$ around $8000$~K (\cite{la98}).

\begin{figure}[t]
  \begin{center}
  \mbox{\includegraphics[width=8.5cm, clip=TRUE]{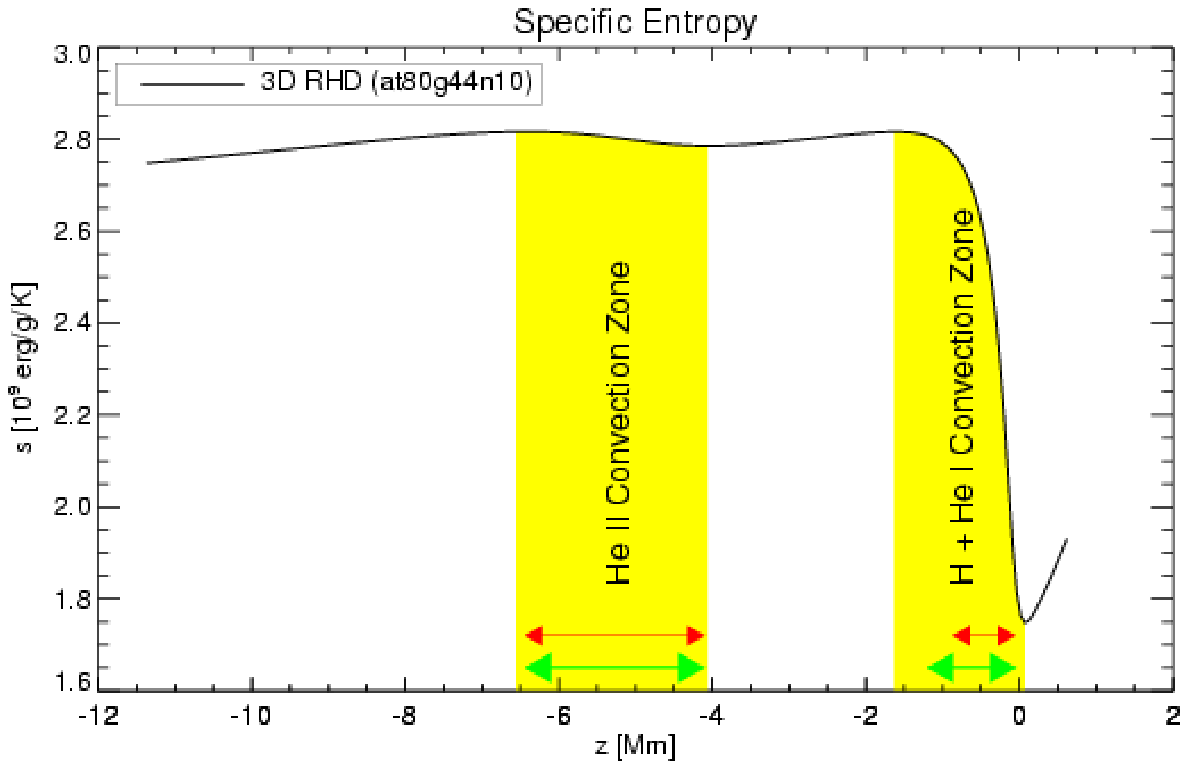}}
  \mbox{\includegraphics[width=8.5cm, clip=TRUE]{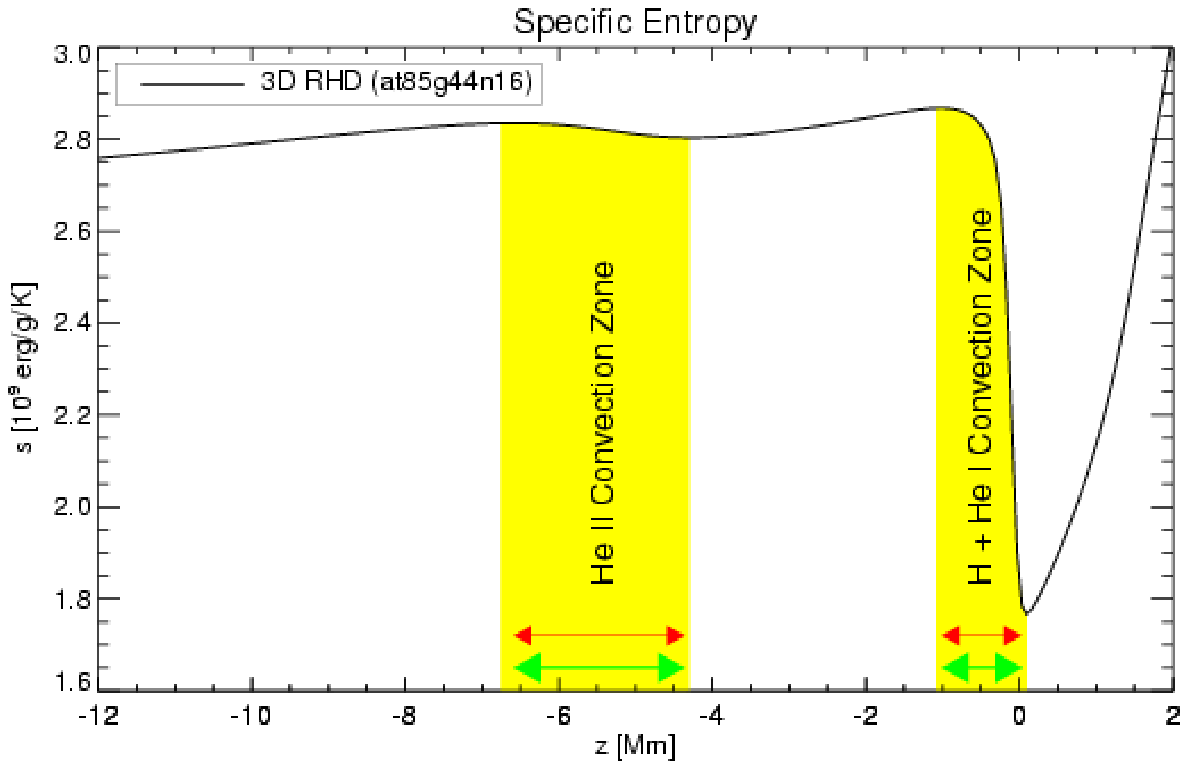}}
\end{center}
\caption{
  \emph{Mean specific entropy} $\langle s(z) \rangle_{x,y,t}$ for Model~1
  (top) and \mbox{Model~2} (bottom).  In both cases,
  the entropy stratification indicates two separate convection zones
  ($ds/dz<0$, shaded). The borders of convective
  instability according to local MLT models are indicated by 
  thin ($\alpha$=$0.5$) and thick ($\alpha$=$1.0$) arrows.
\label{fig2}}
\end{figure}

\begin{figure}[htb]
  \begin{center}
  \mbox{\includegraphics[width=8.5cm]{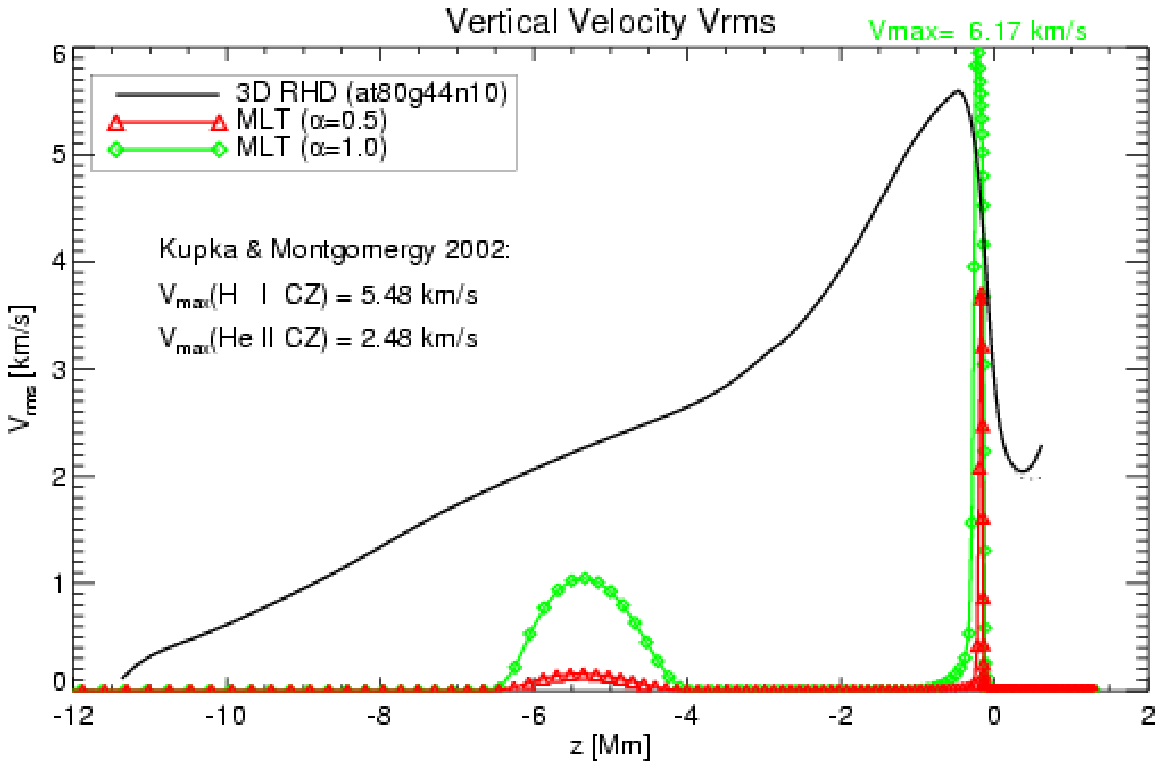}}
  \mbox{\includegraphics[width=8.5cm]{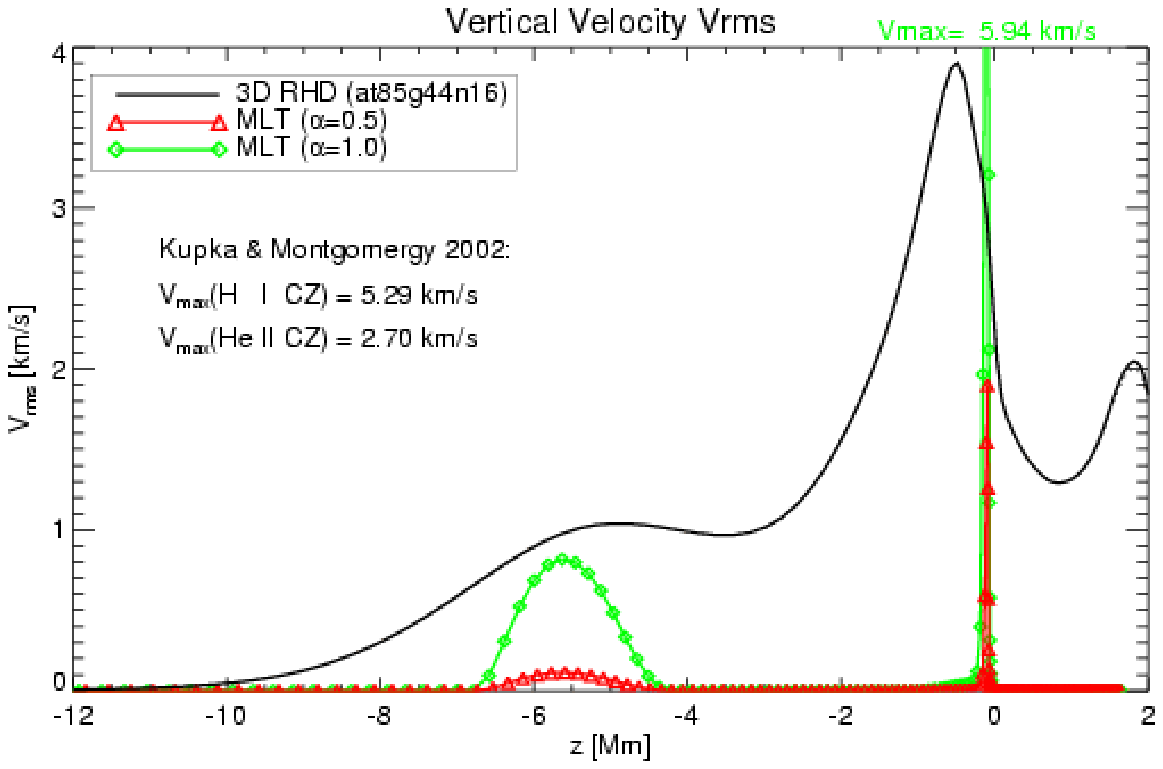}}
\end{center}
\caption{
\emph{Mean vertical velocity} $\sqrt{\langle V_{\rm z}^2(z) \rangle_{x,y,t}}$,
compared with MLT models. According to 3D hydrodynamics,
both convection zones are connected by overshooting flows. The extended
exponentially decaying flow field below the He\,II convection zone is also
a result of 'overshoot'. 
\label{fig3}}
\end{figure}

\begin{figure}[htb]
  \begin{center}
  \mbox{\includegraphics[width=8.5cm]{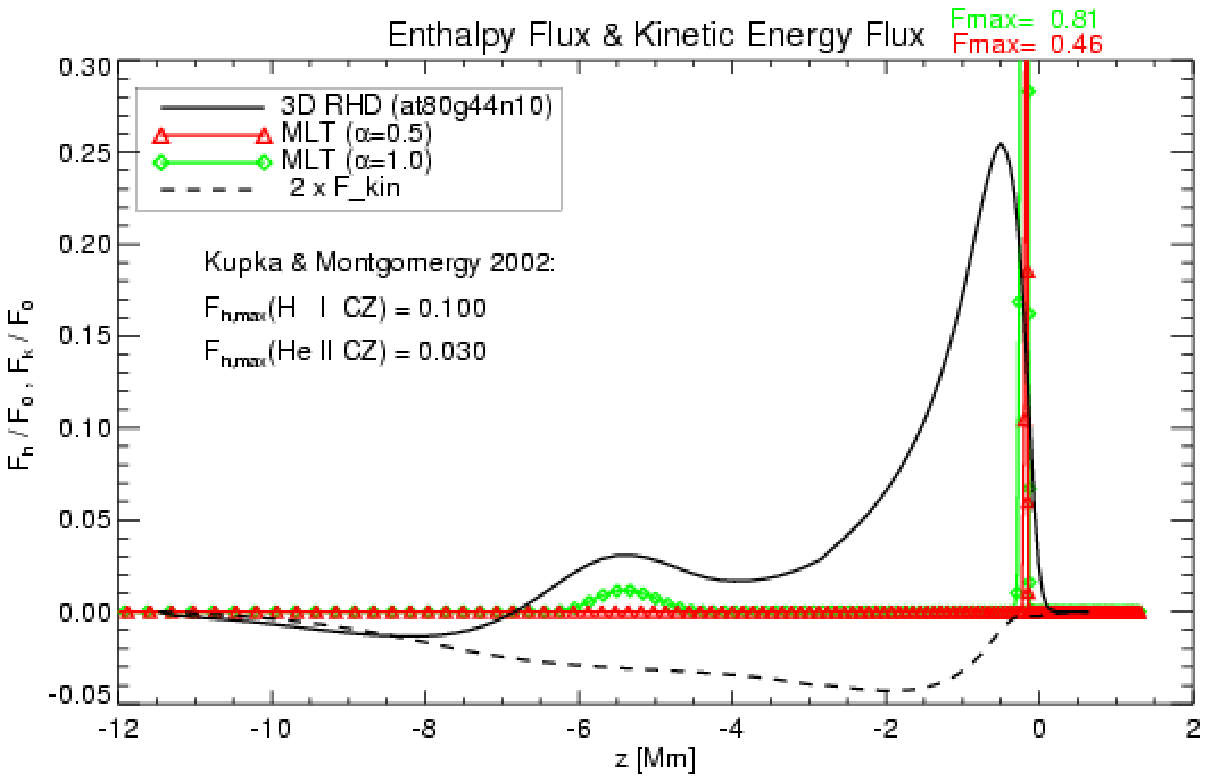}}
  \mbox{\includegraphics[width=8.5cm]{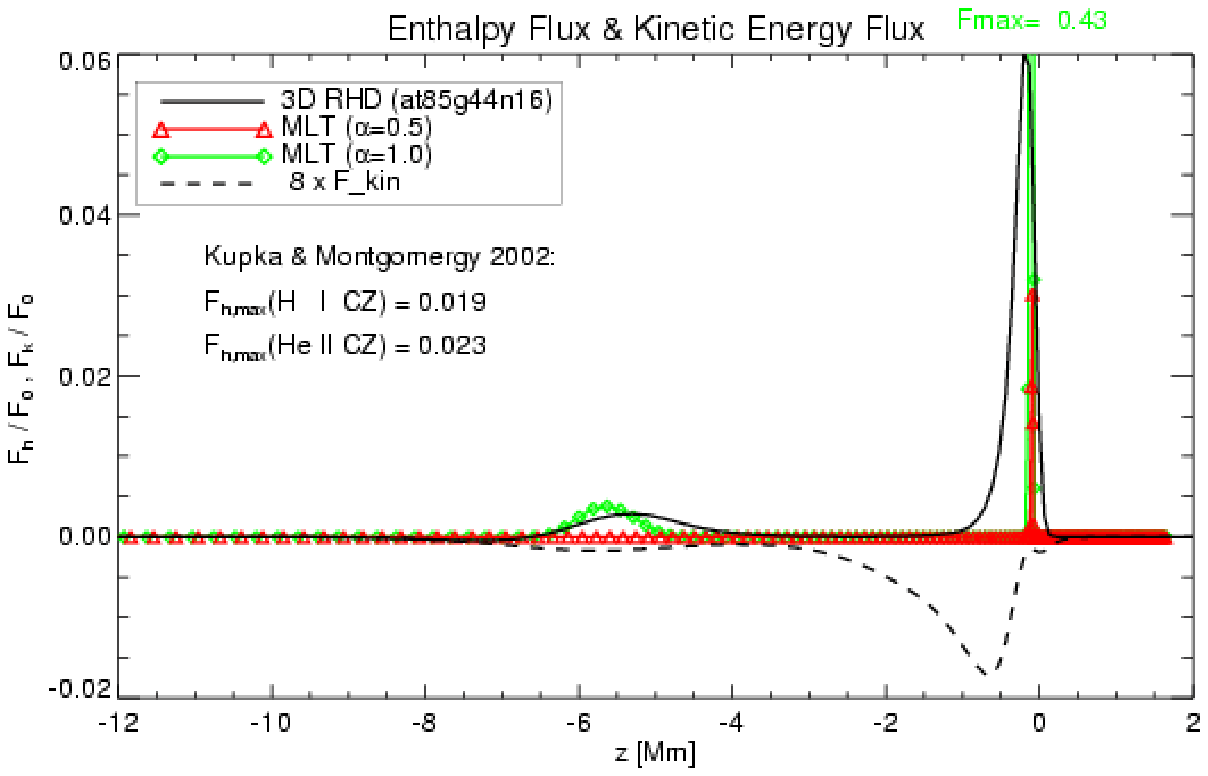}}
\end{center}
\caption{
\emph{Mean enthalpy flux} $\langle \rho V_{\rm z} \, h \rangle_{x,y,t}$, 
compared with mixing-length results. Like in the solar granulation,
the 3D simulations give a \emph{mean kinetic energy flux} 
$\langle \rho V_{\rm z} \, \vec{V}^2/2 \rangle_{x,y,t}$
which is directed downwards at all heights.
\label{fig4}}
\end{figure}

\section{3D Hydrodynamical Convection Models}
\label{sec:3DRHD}

The numerical simulations presented here were performed with
\mbox{CO$^5$BOLD}, a 3D radiation hydrodynamics code designed to model
stellar convection (see \cite{fsd02} or \cite{wfslh04} for details).
The integration of the equations of hydrodynamics
is based on a conservative \mbox{finite} volume approach using an
approximate Riemann solver of Roe type together with a \emph{van Leer}
reconstruction scheme. The Roe solver was modified to handle an
external gravity field and an arbitrary tabulated equation of state
(EOS). For the present application, we use a realistic EOS table
accounting for partial ionization of hydrogen and helium (as well as
H$_2$ molecule formation).

The 3D non-local radiative transfer is solved on a system of
\emph{long} rays, employing a modified Feautrier \mbox{scheme}. 
Using a realistic Phoenix-OPAL Rosseland mean opacity table, 
we adopt the grey approximation in this exploratory study. Strict 
LTE is assumed (no scattering), and radiation pressure is ignored.

The simulations are performed on a \mbox{Cartesian} grid with variable
cell size in the vertical direction. We apply periodic lateral
boundary conditions, while top and bottom boundaries are 'closed'.

\section{Results}
\label{sec:results}

The results shown here are derived from convection simulations for
stellar parameters $T_{\rm eff}=8000$~K, $\log g=4.4$
(model~1), and $T_{\rm eff}=8500$~K, $\log g=4.4$ (model~2),
respectively.  A representative snapshot from each of these sequences
is displayed in Fig.~\ref{fig1}. Similar calculations have been performed
for $\log g=4.0$, but are not shown here. Further details can be found in 
\cite*{fs04}.

\subsection{Vertical Structure, comparison with MLT}
\label{sec:vertical}

\begin{table}[hbt]
  \caption{Comparison of maximum convective velocity, $Vc_{\rm max}$
           {\upshape ([km/s])}, maximum fraction of convective energy 
           flux, $Fc_{\rm max}$,  and of kinetic energy flux 
           ($Fk_{\rm max}\equiv \max(|Fk|)/F$) for different 
           kinds of A-type star convection 
           models. (u) and (l) refer to the upper (H+He\,I) 
           and lower (He\,II) convection zone, respectively.}
  \label{tab1}
  \begin{center}
    \leavevmode
    \footnotesize
    \begin{tabular}[h]{lcccc}
      \hline \\[-5pt]
       & Hydrodyn.         & MLT          &  MLT         &   KM02   \\
       & Simulation        & $\alpha=0.5$ & $\alpha=1.0$ &   Theory \\[+5pt]
      \hline \\[-5pt]
      $T_{\rm eff} = 8000$~K, & $\log g = 4.40$: &  &  &  \\
      \hline \\[-4pt]
      $Vc_{\rm max}$(u) & 5.59   & 3.71   & 6.17   & 5.48   \\
      $Vc_{\rm max}$(l) & -----  & 0.16   & 1.05   & 2.48   \\
      $Fc_{\rm max}$(u) & 0.255  & 0.458  & 0.806  & 0.100  \\
      $Fc_{\rm max}$(l) & 0.031  & 0.000  & 0.012  & 0.030  \\
      $Fk_{\rm max}$(u) & 0.0215 & -----  & -----  & 0.0011 \\
      $Fk_{\rm max}$(l) & ------ & -----  & -----  & 0.0012 \\
      \hline \\[-5pt]
      $T_{\rm eff} = 8500$~K, & $\log g = 4.40$: &  &  &  \\
      \hline \\[-4pt]
      $Vc_{\rm max}$(u) & 3.90   & 1.90   & 5.94   & 5.29   \\
      $Vc_{\rm max}$(l) & 1.04   & 0.11   & 0.82   & 2.70   \\
      $Fc_{\rm max}$(u) & 0.060  & 0.030  & 0.430  & 0.019  \\
      $Fc_{\rm max}$(l) & 0.003  & 0.000  & 0.004  & 0.023  \\
      $Fk_{\rm max}$(u) & 0.0021 & -----  & -----  & 0.0004 \\
      $Fk_{\rm max}$(l) & 0.0002 & -----  & -----  & 0.0010 \\
      \hline \\[-15pt]
      \end{tabular}
  \end{center}
\end{table}

In \mbox{Figs.~\ref{fig2} - \ref{fig4}}, the mean vertical structure
of the 3D hydrodynamical simulations, obtained by horizontal and
temporal averaging, is compared with the results of standard MLT, in
the version described by \cite*{mih78}, for mixing length parameters
$\alpha$=$0.5$ and $1.0$. Clearly, \emph{MLT does not even approximately 
match the hydrodynamical results}, no matter what value of $\alpha$ is
chosen. We note that the non-local convection models by \cite*{km02}
are qualitatively more similar to the hydrodynamical solutions,
although considerable differences remain, especially in the energy
fluxes. In contrast to the findings by KM02, our 3D models do not show
anywhere a positive kinetic energy flux. Table \ref{tab1} lists a some
key numbers characterizing the different kinds of models.

Fig.~\ref{fig4} demonstrates that 'overshoot' below the He\,II
convection zone is substantial. The exponential tail of the velocity
field is clearly seen in model~2, where the velocity scale height in
terms of the pressure scale height at the bottom of the He\,II CZ is
$H_v/H_p \approx 0.4$. Model~1 is not deep enough to include the
exponential part of the overshoot region. We estimate {$H_v/H_p \la 0.7$}.

\subsection{Horizontal Structure}
\label{sec:horizontal}

The horizontal structure emerging from our 3D simulations of surface
convection in A-type stars is evident from the intensity images
displayed in Fig.~\ref{fig5}. Obviously, the flow topology is
qualitatively similar to that of the solar granulation, i.e. isolated
hot up-flows (granules) are separated by a network of connected cool
down-flows (intergranular lanes). Due to a more efficient radiative
energy exchange, the granules at the surface of A-type stars are
relatively larger than on the Sun, and seem to show less
sub-structure. In fact, the filling factor of dark areas (where $I <
\overline{I}$) is $f_d \approx 0.34$ for both model~1 and model~2,
compared to $f_d \approx 0.53$ for the Sun.

\subsection{Synthetic Line Profiles}
\label{sec:linfor}

\begin{figure*}[t]
  \begin{center}\mbox{
  \includegraphics[width=5.5cm, clip=true]{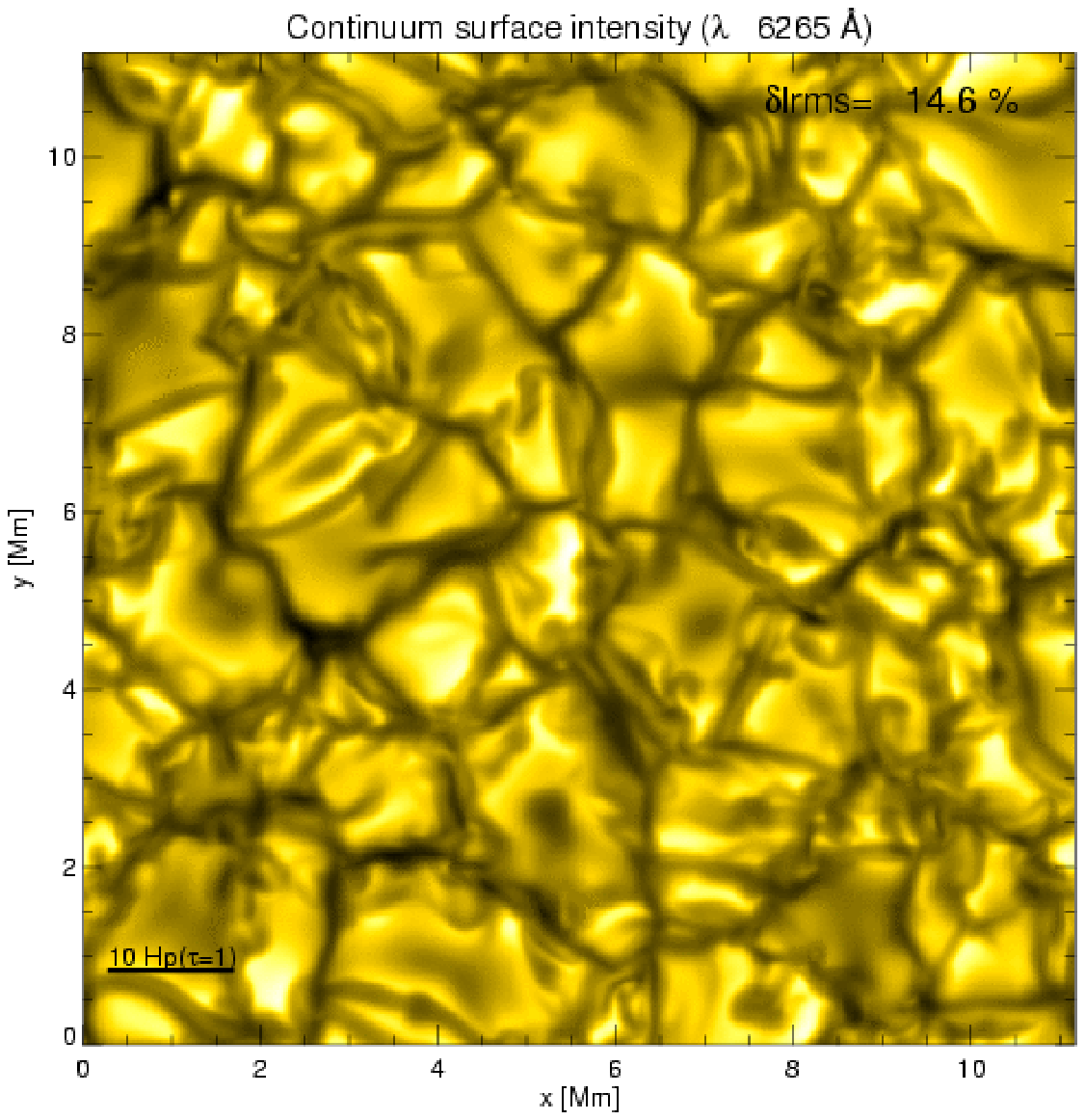}
  \includegraphics[width=5.5cm, clip=true]{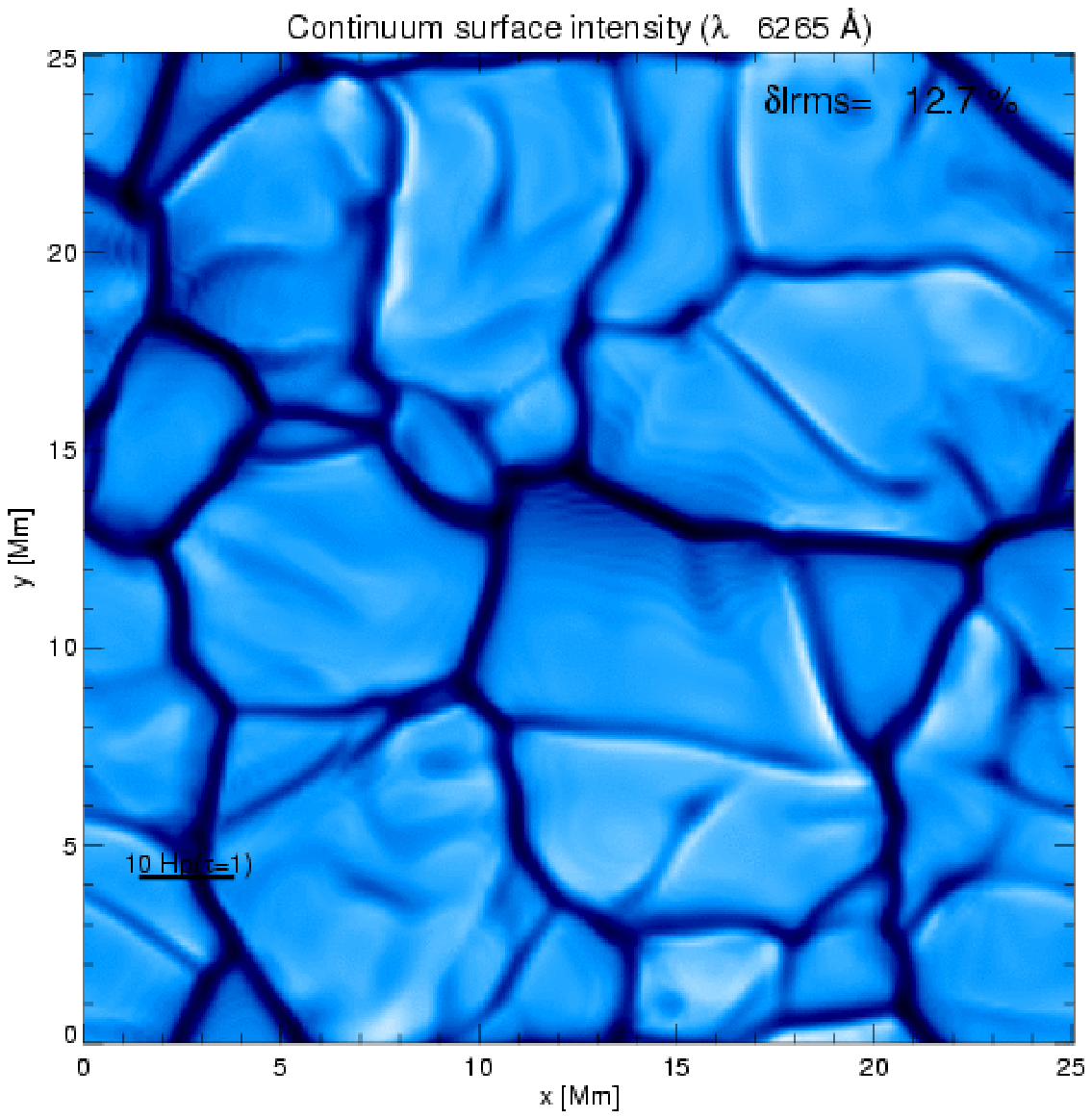}
  \includegraphics[width=5.5cm, clip=true]{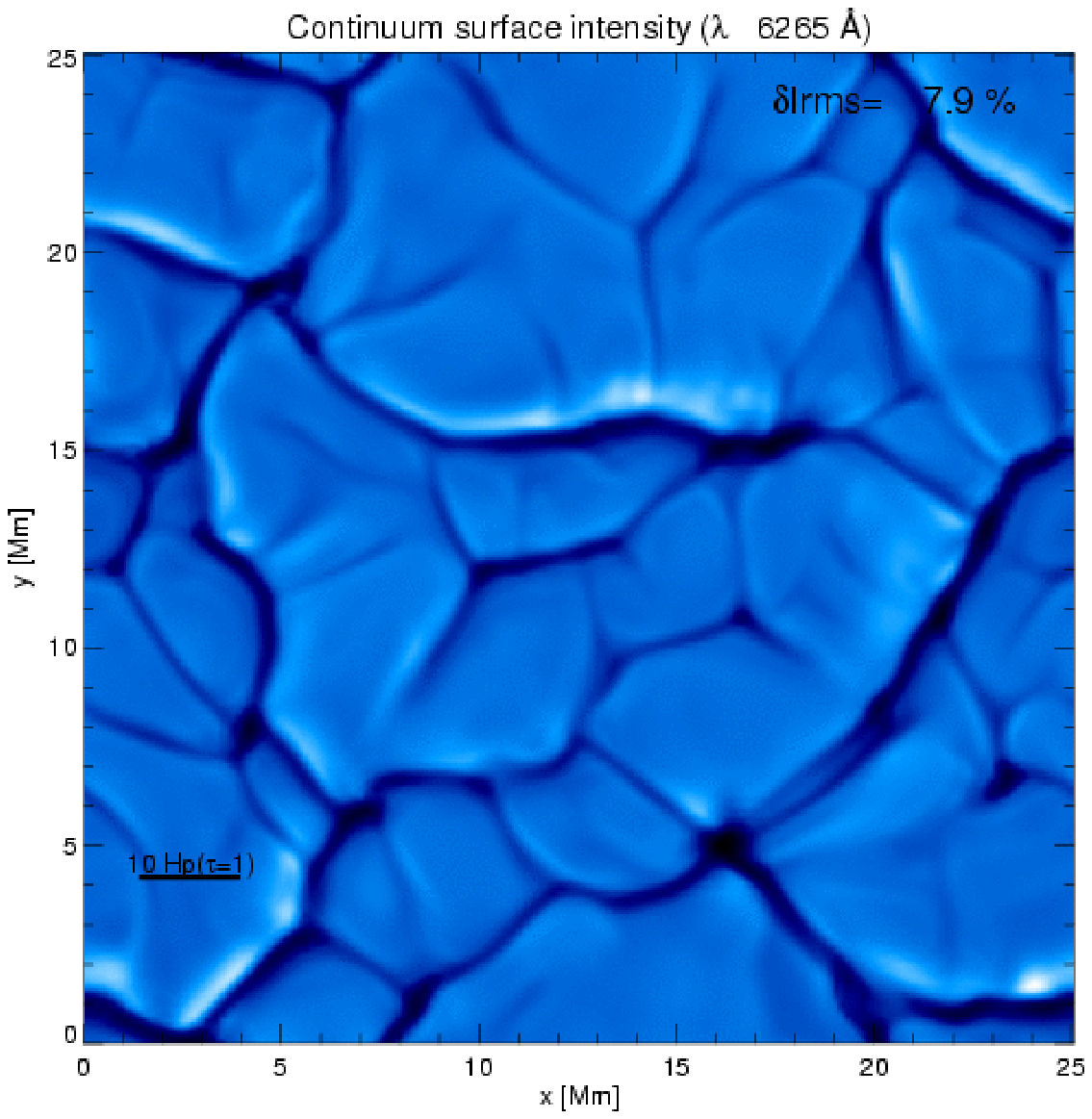}}
\end{center}
\caption{
Emergent continuum intensity at $\lambda\, 6265$\,\AA\ resulting from
3D hydrodynamical simulations of the solar granulation
{\bf (left, $\delta I_{\rm rms}$=$14.6\%$)} and of surface convection 
in main-sequence A-type stars with 
$T_{\rm eff}=8000$~K {\bf (middle, $\delta I_{\rm rms}$=$12.7\%$)} and
$T_{\rm eff}=8500$~K {\bf (right,  $\delta I_{\rm rms}$=$ 7.9\%$)}.
\label{fig5}}
\end{figure*}

\begin{figure*}[htb]
  \begin{center}\mbox{
  \includegraphics[width=6.0cm, clip=true]{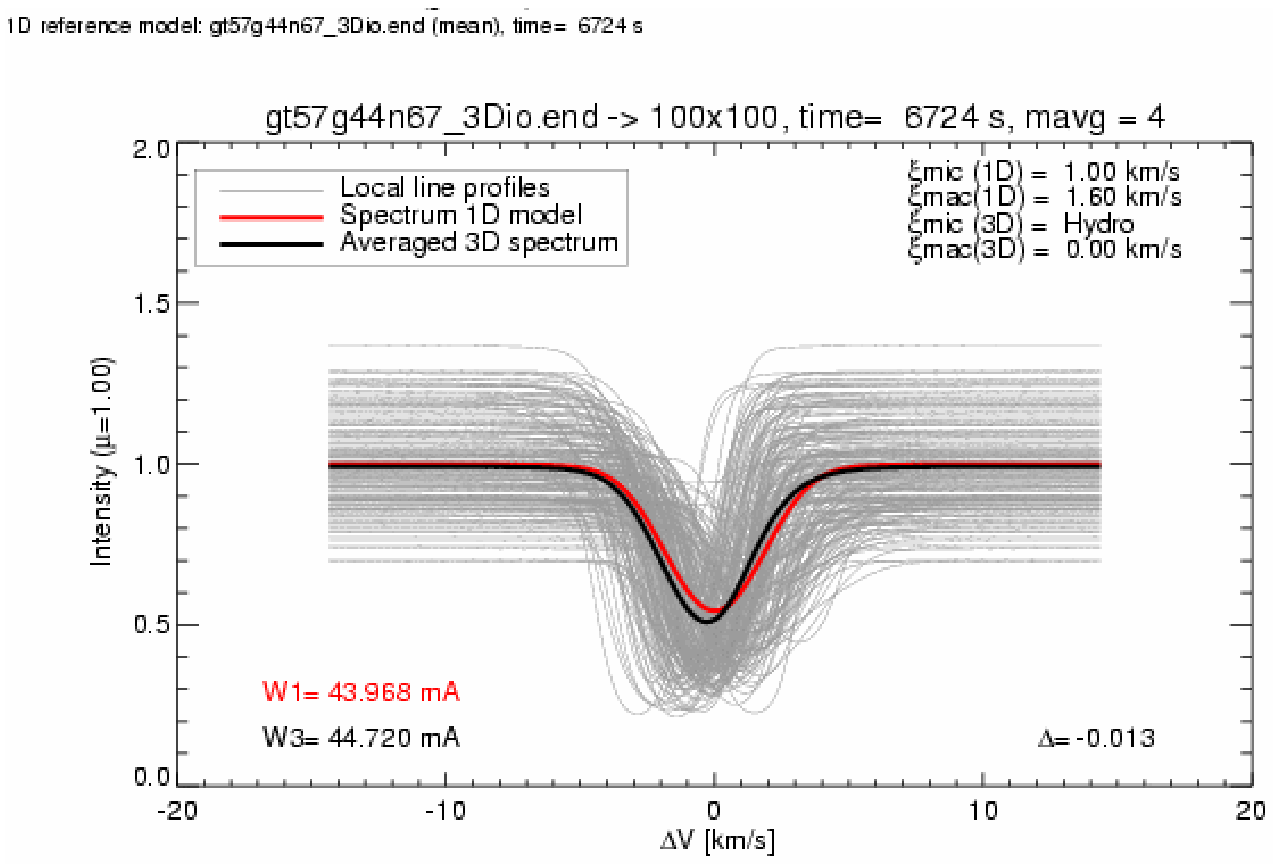}
  \includegraphics[width=6.0cm, clip=true]{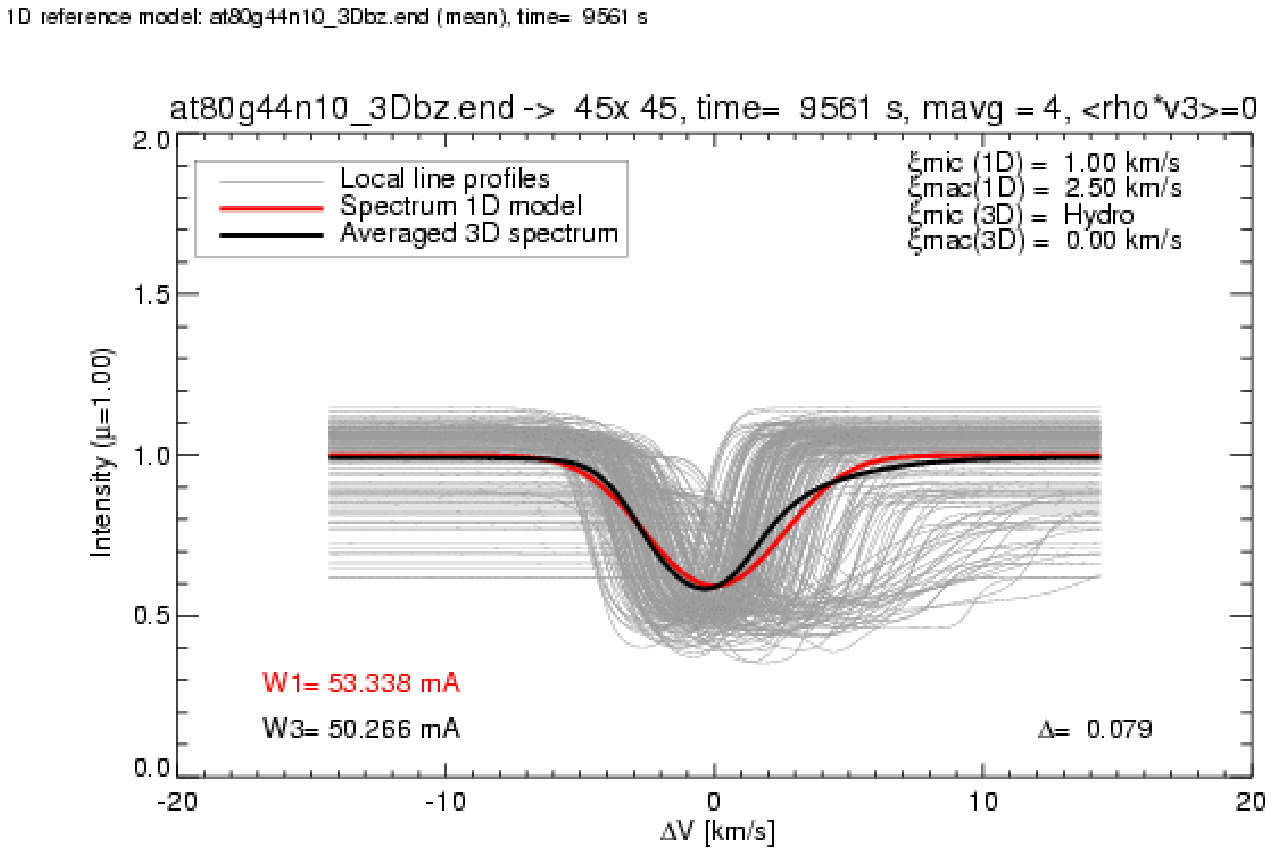}
  \includegraphics[width=6.0cm, clip=true]{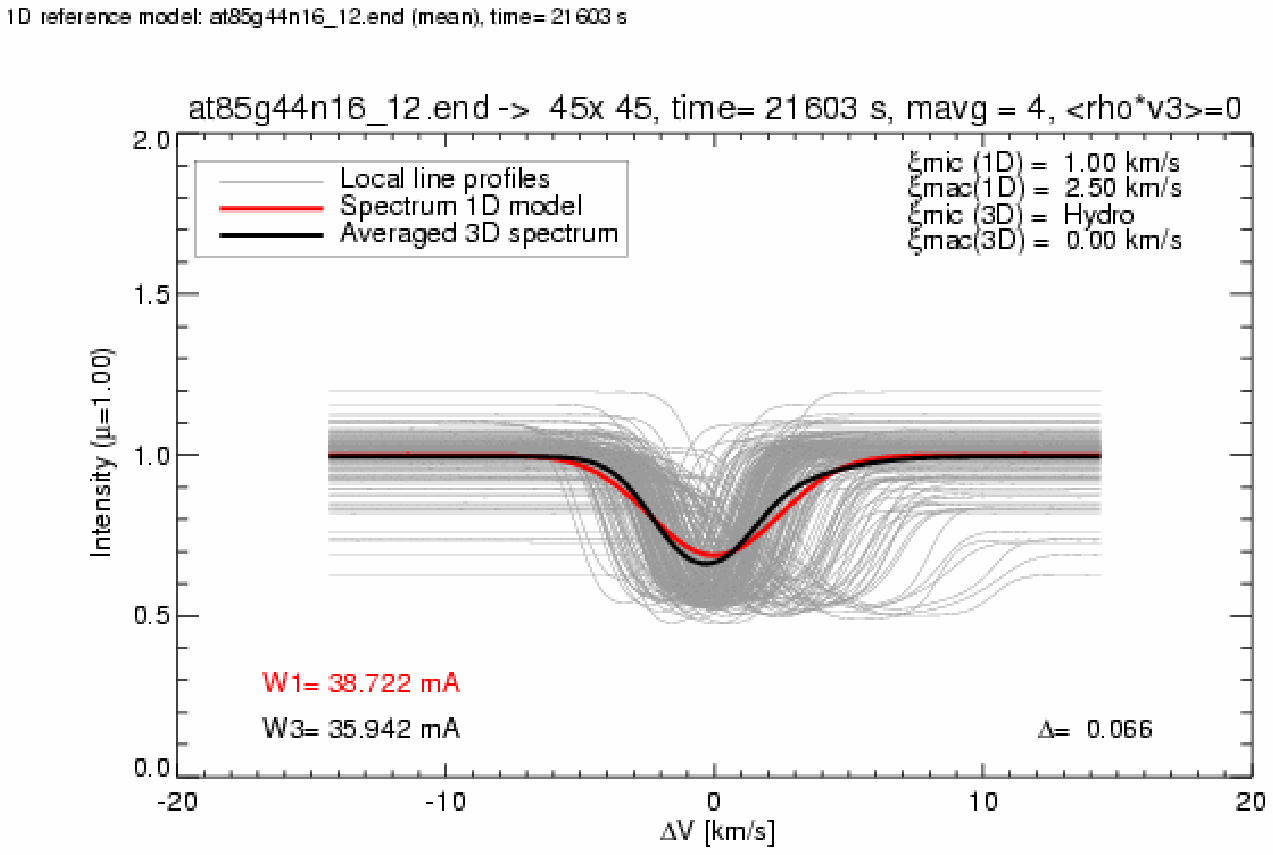}}\\
\mbox{
  \includegraphics[width=6.0cm, clip=true]{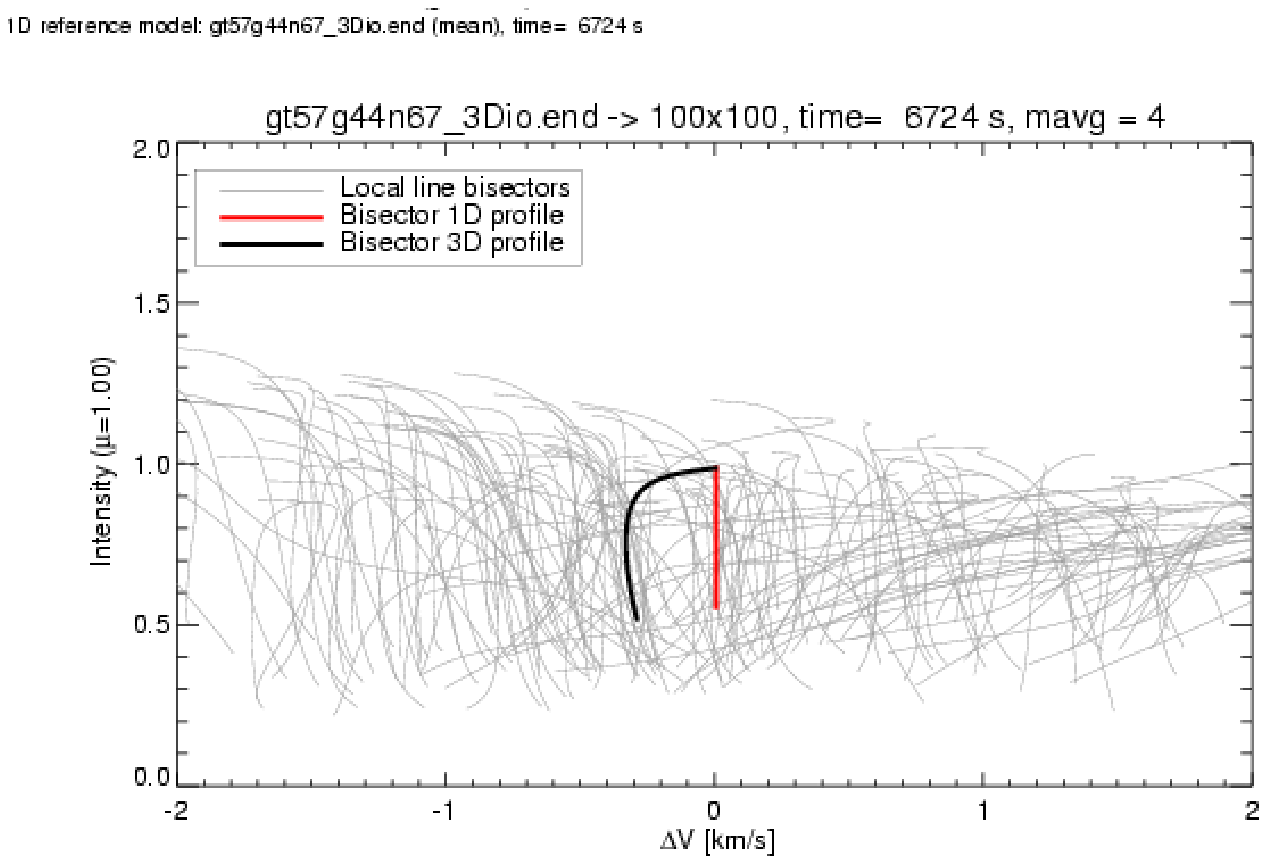}
  \includegraphics[width=6.0cm, clip=true]{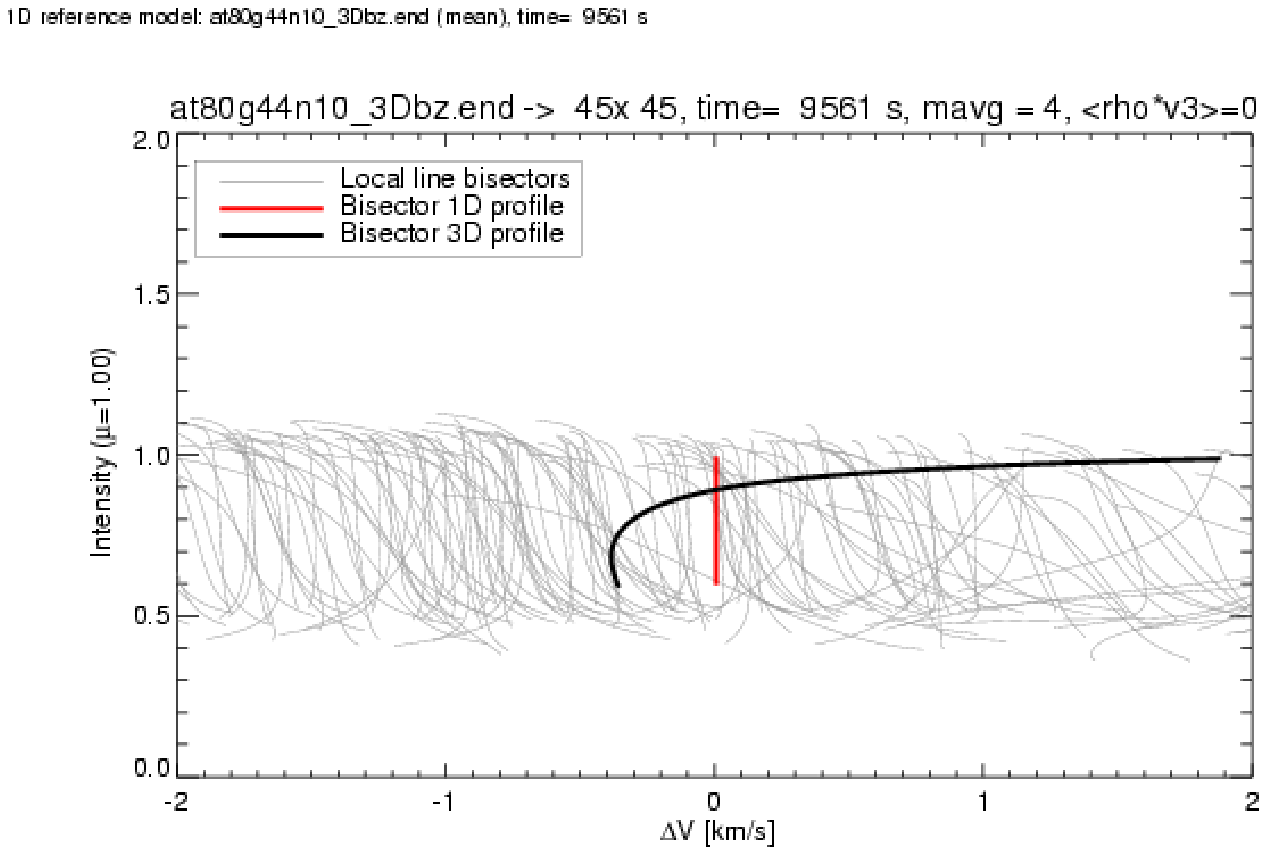}
  \includegraphics[width=6.0cm, clip=true]{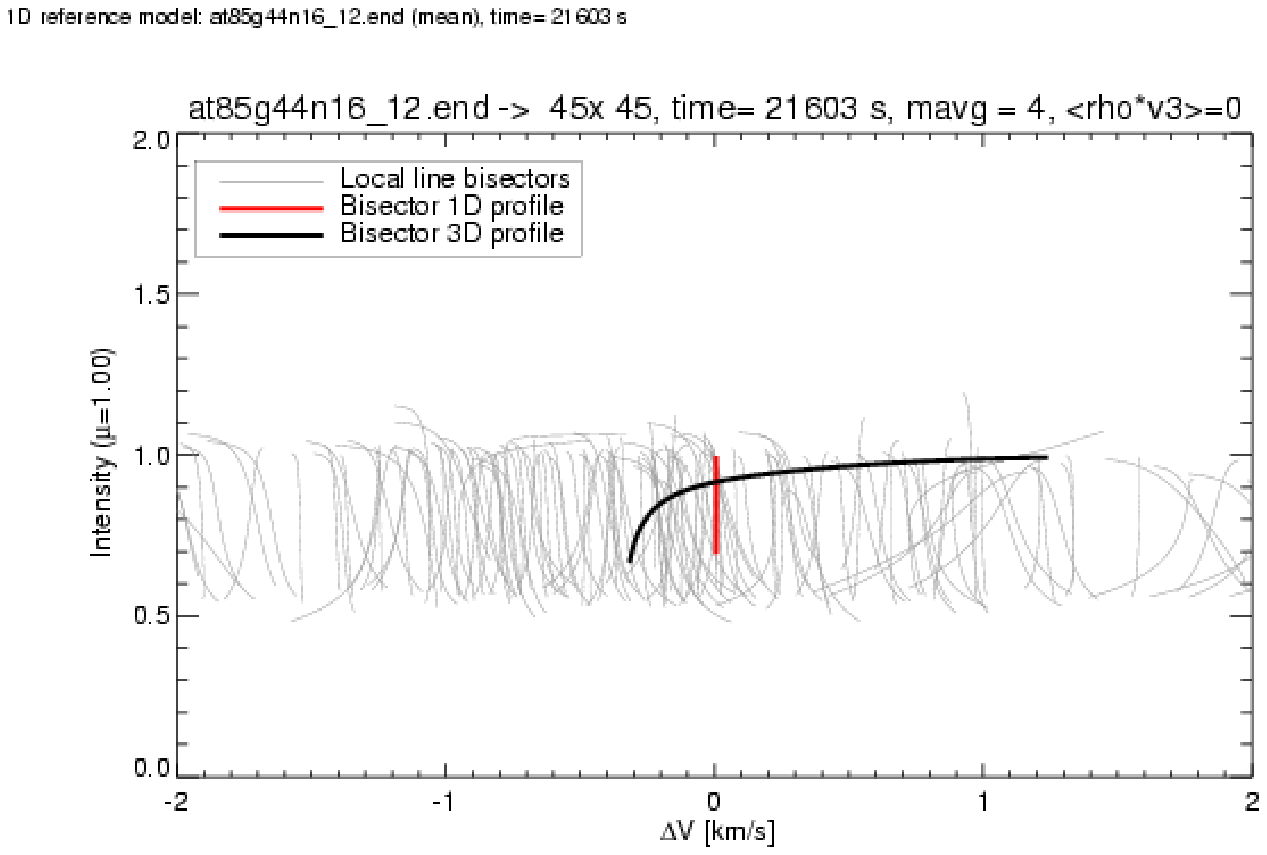}}
  \end{center}
\caption{
Spatially resolved and averaged line profiles ({\bf top})
and line bisectors ({\bf \mbox{bottom}}) of Fe\,I 
$\lambda\, 6265.13$\,\AA, computed from the snapshots shown in 
Fig.~5 for vertical lines-of-sight ($\mu=1$). 
For the Sun, the $gf$-value has been reduced by a factor 100.
The asymmetry of the \emph{flux profiles} (not shown) is 
qualitatively similar.
\label{fig6}}
\end{figure*}

For the snapshots shown in Fig.~\ref{fig5}, we have computed synthetic
line profiles both for vertical rays (disk-center) and for integrated
light (flux) under the assumption of LTE. The resulting disk-center
line profiles and line bisectors of Fe\,I $\lambda\, 6265.13$\,\AA\ are
presented in Fig.~\ref{fig6}. The considerable photospheric velocities
and temperature fluctuations induce a distinct asymmetry of the
emergent line profiles.  We have investigated several snapshots and
different spectral lines, and found that the line bisector always
exhibits a solar-like C-shape, but with a larger excursion to the red
near the continuum. In the flux spectra, the line bisectors span
typically 2~km/s and 1~km/s for models~1 and 2, respectively. This is
of the same order of magnitude as observed by \cite*{la98}, but the
asymmetry is in the opposite direction. The C-shape persists for
the simulations with $\log g$=$4.0$.

\section{Conclusions}
\label{sec:concl}

  The analysis of our 3D hydrodynamical \mbox{simulations}
  indicates a severe failure of the standard local
  mixing-length theory in the regime of A-type star shallow surface
  convection.  The non-local convection
  model by \cite*{km02} gives a much better description of the velocity
  field, but alarming differences remain in the energy fluxes. 
  Overshoot below the He\,II convection zone is found to be substantial,
  in basic agreement with KM02.
  
  According to the simulations, the granulation pattern forming at the
  surface of A-type stars has a solar-like flow topology, with
  granules that are relatively larger than on the Sun and seem to show
  less sub-structure.  Synthetic LTE line profiles based on the
  current 3D convection models of A-type stellar atmospheres show a
  depressed \emph{red} wing, in apparent contradiction to the observed 
  line asymmetry (\cite{la98}). 
  A possible reason for this discrepancy might be missing physics in the 
  simulations (e.g. magnetic fields).
  On the other hand, we note that the slowly rotating A-type stars with
  $T_{\rm eff} \approx 8000$~K observed by \cite*{la98} are classified 
  as Am type and known to be spectroscopic binaries; hence they may be 
  peculiar.

%
%

%
%

\end{document}